\documentclass[doublecol]{epl2}
\usepackage{bm,amsmath,epsfig}
\usepackage{color}

\newcommand{\tw}{t_{\mathrm{w}}}
\newcommand{\tm}{\tau_{\mathrm{m}}}

\newcommand{\tr}{\tau_{\mathrm{\alpha}}}

\title{Density controls the kinetic stability of ultrastable glasses}

\author{Christopher J. Fullerton \and Ludovic Berthier}

\institute{Laboratoire Charles Coulomb, 
UMR 5221 CNRS \& Universit\'e de Montpellier, 34095 Montpellier, France}


\abstract{We use a swap Monte Carlo algorithm to numerically prepare 
bulk glasses with kinetic stability comparable to that 
of glass films produced experimentally by physical
vapor deposition. By melting these systems into the liquid
state, we show that some of our glasses retain their amorphous 
structures longer than $10^5$ times the equilibrium 
structural relaxation time. This 
`exceptional' kinetic stability cannot be 
achieved for bulk glasses produced by slow cooling.
We perform simulations at both constant volume and constant 
pressure to demonstrate that the density mismatch 
between the ultrastable glass and the equilibrium liquid 
accounts for a major part of the observed kinetic stability.}

\begin{document}

\maketitle

{\it Introduction ---} 
Physical vapor deposition is an efficient way
to prepare amorphous thin films with tunable physical properties. 
Molecules are slowly deposited onto a substrate held 
at constant temperature, and a glassy film is 
constructed layer by layer~\cite{BE_2016}.
For well chosen substrate temperatures, the 
resulting glass may exhibit `exceptional'~\cite{swallen_2007} physical properties. 
It can have higher density~\cite{dalal_2012,dalal_2012_2}, 
lower enthalpy~\cite{kearns_2008,ramos_2011} and lower heat 
capacity~\cite{kearns_2010,ahrenberg_2012} than glasses conventionally 
prepared by slow cooling.
These vapor-deposited glasses 
have been classified as `ultrastable', and 
have now been prepared from a wide range of 
molecules~\cite{whitaker_2013,tylinski_2015,walters_2015,viejo2016,toto_2016}.
Although produced in an unusual way, these glasses are thought
to be equivalent to glasses that have been aged for unacheivably long times.

The kinetic stability of vapor deposited glasses 
can be estimated in two ways, both of which involve melting the glass.
A glass can be heated slowly and the `onset temperature' 
at which it starts to melt back to the liquid 
state measured~\cite{swallen_2007}.
The higher the onset temperature, the more stable the glass.
The second measure is through a `stability ratio' which
allows direct comparison of glasses formed 
from different materials~\cite{sepulveda_2014}.
The material is rapidly heated above the glass transition, 
and the ratio between the time it takes the glass to melt and
the equilibrium relaxation time at the melting
temperature is measured. This is the stability ratio, ${\cal S}$.
For vapor-deposited ultrastable glasses, the stability ratio 
is found in the range ${\cal S}=10^2$ (for materials with
low stability~\cite{toto_2016})
to $10^4$ (for the majority of ultrastable glasses), up to
$10^5$ for trisnaphthylbenzene~\cite{sepulveda_2014} and $10^{5.2}$ for $o$-terphenyl~\cite{whitaker_2015},
which seems to set the experimental record.

Ultrastable glasses represent a new class of amorphous materials
with interesting applications~\cite{BE_2016}, but their 
properties are not well understood yet. For instance, 
it is not known how to quantitatively 
relate the degree of equilibration of 
ultrastable glasses to their measured kinetic stability, 
despite recent progress in this 
direction~\cite{wolynes_2009,Wisi_2013,Jack_2016}.  
Computer simulations provide a valuable tool for achieving this understanding, 
as complete 
knowledge of microscopic information provides direct insight into 
the properties of stable glasses. However, 
computational work in this area is challenging, as materials this stable
have effective preparation times that are extremely large.  
Several efforts have been made to simulate 
stable glasses using very slow cooling~\cite{staley_2015,staley_2016},
random pinning~\cite{hocky_2014}, 
nonequilibrium sampling~\cite{jack_when},
or by directly simulating the deposition 
process~\cite{singh_2013,lyubimov_2013,helfferich_2016,flenner17}, 
but the largest reported stability ratio to date remains a
modest ${\cal S} \sim 10^2$~\cite{staley_2015}.
In this article, we report stability ratios that 
can be as large as ${\cal S} \approx 10^5$ for a simulated 
bulk glass-former, comparing favourably with the 
largest values reported in experiments for ultrastable glassy films.
We achieve this record value by preparing glasses 
using swap Monte Carlo~\cite{berthier_2016_swap,NBC17}. 
By considering how these glasses melt in different 
numerical ensembles (isochoric or isobaric), we demonstrate that a major 
part of their large kinetic stability stems from the density 
mismatch between the ultrastable glass and the equilibrium fluid, 
because the dense glass needs to expand to accomodate the 
invading fluid during melting.

{\it Model and simulations ---} 
We study systems of polydisperse hard spheres in three dimensions.
The spheres have a continuous distribution of diameters, 
$P(\sigma_{\mathrm{min}} \leq \sigma \leq \sigma_{\mathrm{max}}) = 
A/\sigma^3$, where $A$ is a normalization constant.
We choose $\sigma_{\mathrm{min}}$ and $\sigma_{\mathrm{max}}$ to give a 
polydispersity of $\Delta = \sqrt{\langle \sigma^2 \rangle - \langle \sigma \rangle^2}/\langle \sigma \rangle = 23 \%$, which ensures the efficiency of the swap 
algorithm while preventing the system from 
crystallising too easily~\cite{berthier_2016_swap}.
The interaction strength between 
particles $i$ and $j$ is infinite if the interparticle 
distance is smaller than $\sigma_{ij}$ and is zero otherwise.
We determine $\sigma_{ij}$ using a nonadditive rule, $\sigma_{ij} = (\sigma_i + 
\sigma_j)(1-\epsilon|\sigma_i-\sigma_j|)/2$.
When $\epsilon = 0$ the particles have a regular
additive hard sphere interaction.
The interaction is non-additive when $\epsilon \neq 0$.
Non-additive hard spheres are less prone to 
crystallisation than additive ones, allowing us to age non-additive glasses for extremely long times.
We melt glasses of three types:
most of our results are for glasses with $N = 1000$ and $\epsilon = 0$, 
but we also present results from systems with $N = 8000$ and $\epsilon = 0$, 
and with $N = 300$ and $\epsilon = 0.2$. The non-additive model is the most promising in terms of kinetic stability, 
but its bulk behaviour has not been explored yet. For this reason, we have focused our study on the additive model studied in Ref.~\cite{berthier_2016_swap}.

The system is characterized by the packing fraction 
$\phi = \pi \rho \langle \sigma^3 \rangle/6$, where 
$\rho = N/V$ is the number density and $\langle 
\sigma^3 \rangle$ is the average of the cube of the particle diameter.
Uniquely for the hard sphere fluid, as compared to 
more generic glass-formers, the temperature $T$ and pressure $P$
cannot be varied independently. Instead they always appear as a ratio 
through the reduced 
pressure $p = P/\rho k_B T$, related to $\phi$ by the 
equation of state $p=p(\phi)$ ($k_B$ is the Boltzmann constant). 
To aid in comparison with experiments, 
we define the (adimensional) `volume', $v = \phi^{-1}$, and take 
$1/p \sim T / P$ as the analog of temperature~\cite{berthier_2009}.
Therefore we prepare hard sphere glasses by increasing the pressure 
(equivalent to cooling), and melt them by 
decreasing the pressure (equivalent to heating).
The hard sphere model is thus fully equivalent to continuous
pair potentials for fluids, even though its 
experimental realisation is usually achieved using colloidal
particles. To mimic experiments, we performed two sets of simulations
where ultrastable glasses are either slowly or suddenly 
decompressed (the analogs of slow or sudden heating).

We use an enhanced swap Monte Carlo algorithm to prepare 
the initial ultrastable glass configurations, but use ordinary 
Monte Carlo simulations~\cite{Berthierkob2007} 
to study the kinetics of their melting.  
In ordinary Monte Carlo simulations, we 
hold either the volume or 
the pressure constant~\cite{Frenkel:2002}.
At constant volume, particle translations are carried out by 
chosing a random particle and then randomly displacing it within a cube of size 
$\delta r_0$ centred on the particle.
These moves are rejected if they lead to an overlap between particles.
At constant pressure, volume moves are carried out with probability $p_V$ and 
translational moves with probability $(1-p_V)$.
In a volume move, the volume of the simulation box is changed by a random 
amount $\delta V$ chosen from the interval $[-\delta V_0,\delta V_0]$. 
Volume moves are rejected if they lead to an overlap, and 
accepted with the appropriate Boltzmann weight~\cite{Frenkel:2002}.
We take $\delta r_0 = 0.1$ and $\delta V_0 = 0.2$ for all systems. For the 
systems with $N = 300$ and $N = 1000$ we take $p_V = 1/N$ and for the system 
with $N = 8000$ we take $p_V = 0.01$. Our time unit represents 
$N$ attempted Monte Carlo moves, and lengths are measured in units 
of the average particle diameter.

In the swap Monte Carlo used to prepare initial states, 
additional particle-swap moves are performed. These reduce the 
equilibration time by many orders of 
magnitude~\cite{berthier_2016_swap,NBC17} and allow the production 
of equilibrium configurations at very large pressures 
(the analog of low temperatures).
Equilibration is ensured by checking that time correlation functions
(in particular density-density correlations) have decayed fully as explained in detail in Ref.~\cite{NBC17}, 
and checking that the pressure lies on the equilibrium equation of state~\cite{berthier_2016_swap}.
The swap algorithm and vapor deposition both
generate configurations using `unusual' dynamics 
that are very efficient in regions where the `physical' dynamics 
would completely fail to thermalise the system. Because of the very slow deposition process, vapor deposition thermalises thin films, while swap Monte Carlo acts on bulk configurations.

\begin{figure}
\onefigure[width = 8.5cm]{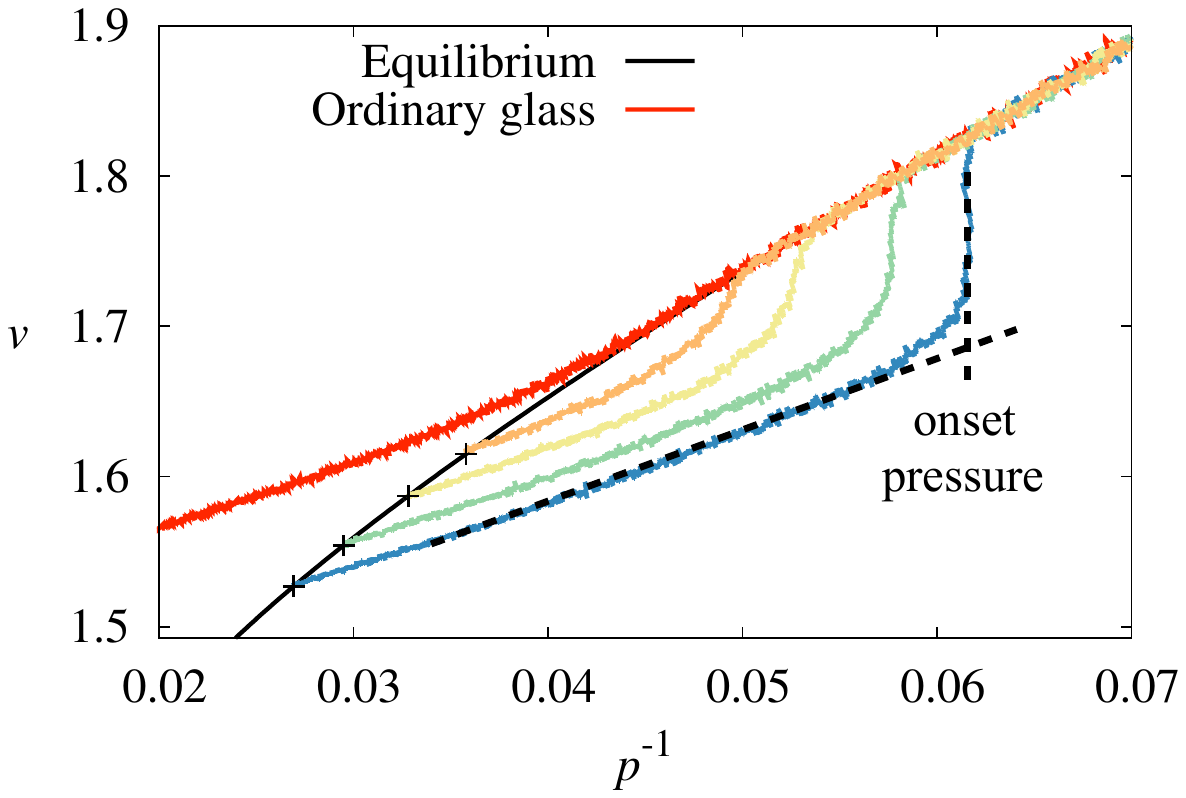}
\caption{Slow decompression of 
hard sphere glasses at constant rate $dP/dt = -10^{-6}$
from various stable initial states (crosses) along the  
equilibrium equation of state (solid black line).
The equation of state of an ordinary glass 
prepared using a slow compression at $dP/dt=10^{-6}$ is also shown.
The crossing point of the two dashed lines allows the onset pressure of 
the most stable glass to be determined.
Stable glasses are (up to 7\%) denser and melt at pressures
(up to 40\%) lower than ordinary glasses.
We quote the decompression rate in unnormalized pressure $P$ (with $k_BT = 1$) for clarity as $dp/dt$ has an additional dependence on $\rho$.
}
\label{fig:slow_decompression}
\end{figure}

{\it Slow melting ---} 
We begin by slowly decompressing 
a selection of ultrastable glasses prepared in various initial 
states. We slowly change the pressure at a constant rate 
and measure the packing fraction $\phi$.  
The results are shown in Fig.~\ref{fig:slow_decompression}.
As each glass is decompressed, its volume increases 
following a nonequilibrium equation of state $v(p)$, which 
describes the expansion of an arrested solid whose 
structure does not relax. 
Each glass follows its own nonequilibrium equation of state until it melts
at a given onset pressure, below which the
system follows the equilibrium equation of state 
of the fluid. 
The onset pressure for a glass can be determined from the crossing point of the two dashed lines shown in Fig.~\ref{fig:slow_decompression}.
More stable glasses are denser, 
and melt at lower pressure, reflecting increasing kinetic stability.
Compared to an ordinary glass slowly cooled through the 
(computer) glass transition, our most stable glasses can be 
denser by about 7~\%, and their onset pressure decreases by about 
40~\%. Similar `exceptional' behaviour has been observed in experiments 
carried out on vapor deposited 
glasses~\cite{swallen_2007,dalal_2012,dalal_2012_2}, although these numbers
are sensitive to the details of the 
thermodynamics of the studied material.
To demonstrate that our most stable systems are truly 
ultrastable in the experimental sense, 
we turn to a more general measure of stability.

\begin{figure}
\onefigure[width = 8.5cm]{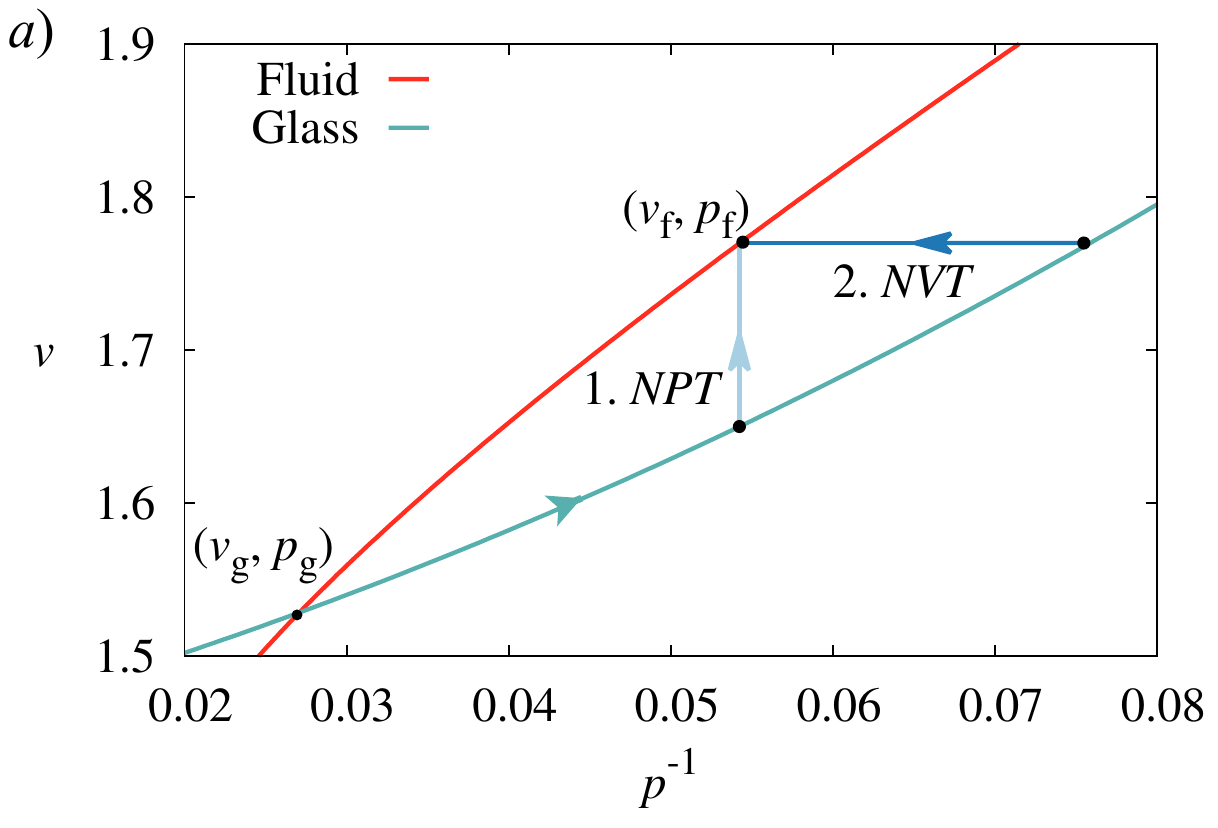}
\onefigure[width = 8.5cm]{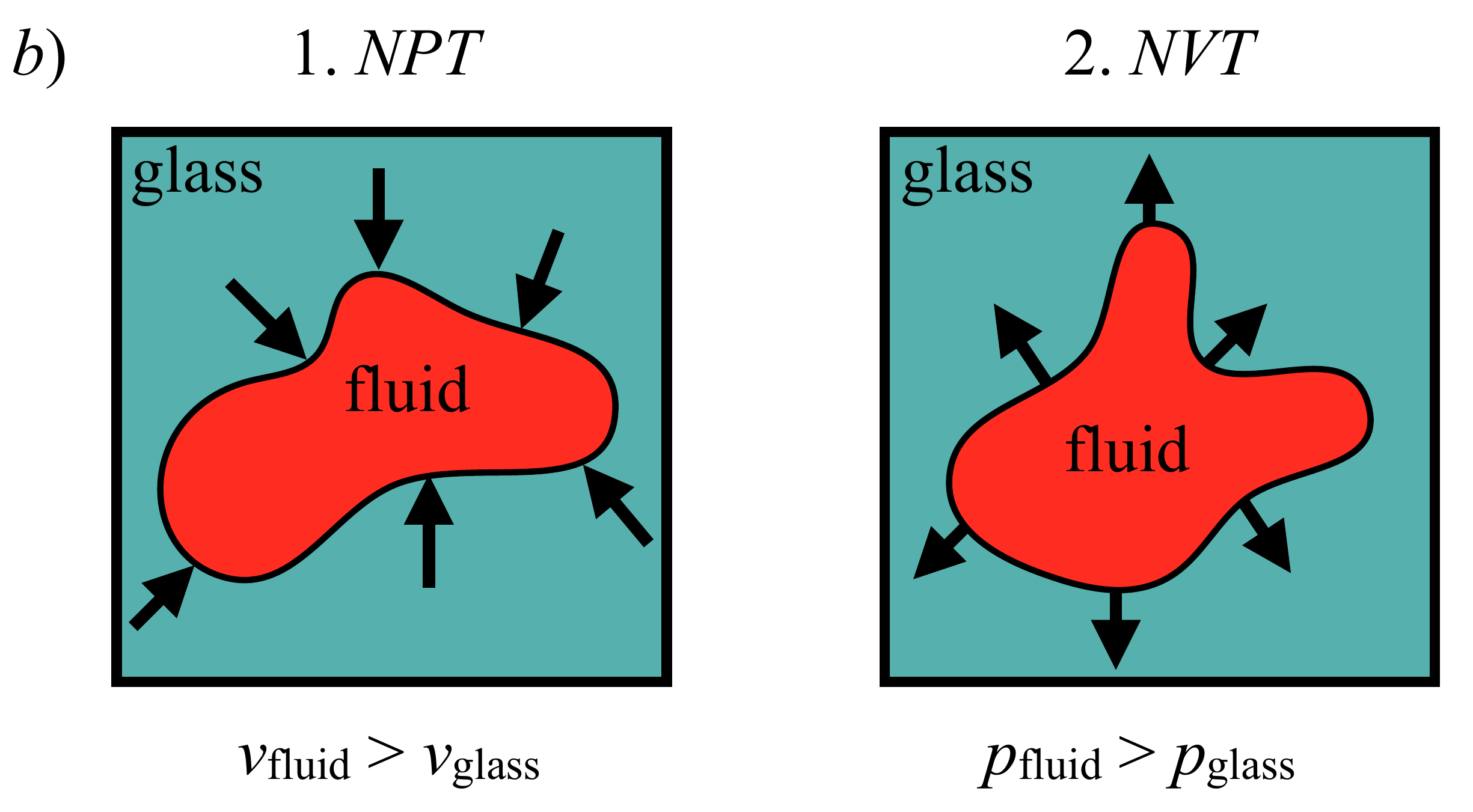}
\caption{a) Melting of a glass prepared at $(v_g,p_g)$ 
to a fluid state at $(v_f, p_f)$ via isobaric 
route 1 or isochoric route 2. In both cases a rapid expansion 
with constant structure along the glass equation 
of state is followed by a slower melting
where either 1) the volume increases or 2) the pressure increases
b) The invasion of the glass by the fluid is 
penalised by the density difference 
(route 1), but is facilitated by the 
pressure difference (route 2).
}
\label{fig:sudden_melt_I}
\end{figure}

{\it Melting at constant volume or pressure ---}
We wish to compare the stability ratio ${\cal S}$
of the simulated glasses with experimental results for
ultrastable glasses. To this end, we 
prepare a glass at a state point characterized by
its volume and pressure, $(v_g, p_g)$, and melt it to 
the fluid at $(v_f, p_f)$, with $v_f > v_g$ and $p_f < p_g$.  
Experiments are performed at constant pressure, but
in simulations we can use either constant volume or constant pressure 
protocols. Although initial and final states are the same,
the kinetics along these two routes are very different,
as illustrated in Fig.~\ref{fig:sudden_melt_I}a. 
In the isobaric case (route 1),
the pressure immediately jumps to the value $p_f$, and the volume
slowly increases towards $v_f$ during melting.  
In the isochoric case (route 2), the system immediately 
jumps to the volume $v_f$, and the pressure slowly increases 
towards $p_f$. 

In both cases, melting begins 
by the appearance of fluid regions within the bulk glass
which slowly invade the entire system, 
as illustrated in Fig.~\ref{fig:sudden_melt_I}b. 
Following the isochoric route 2, the fluid pocket has a larger pressure than 
the glass. These melted fluid regions thus 
push inside the unmelted glass, accelerating the fluid invasion.
Alternatively, following the isobaric route 1, 
the fluid regions are less dense than the glass which 
needs to expand to give way to the fluid. The mechanical work needed
for this expansion penalises the growth of the fluid regions.
Our simulations indicate that the stability ratio of ordinarily-cooled
glasses melted via routes 1 and 2 are comparable (${\cal S} \sim 10^2$)
because these glasses are not dense enough for the above mechanism
to play any role, in agreement with recent 
simulations~\cite{staley_2016}. 
Using stable glasses as initial configurations, we observe that 
the stability ratio measured via route 2 
remains around ${\cal S} \sim 10^2$.
However, it can increase up to ${\cal S} \sim 10^5$ via the
experimentally relevant isobaric route 1 for the same initial and final 
states. This directly demonstrates that the density 
difference between glass and fluid states stabilizes 
dense glasses, and 
that density plays a major role in the `exceptional' kinetic stability 
observed experimentally in vapor-deposited glasses.

This difference in behaviour between ensembles
should occur in non-hard-sphere glasses, where temperature and pressure can be varied independently.
It would be seen if temperature is increased while holding either pressure or volume constant.
A construction similar to Fig. \ref{fig:sudden_melt_I}a can be made in that case, with $T$ instead of $p^{-1}$,
and $V$ or $P$ instead of $v$ (for isobaric or isochoric ensembles respectively).
At constant volume a pressure difference again accelerates melting while 
at constant pressure the density difference stabilises the glass, suggesting our results apply generally to any type of glass-former. 
This claim is confirmed by melting simulations we are currently carrying out on Lennard-Jones glasses.

\begin{figure}
\onefigure[width = 8.5cm]{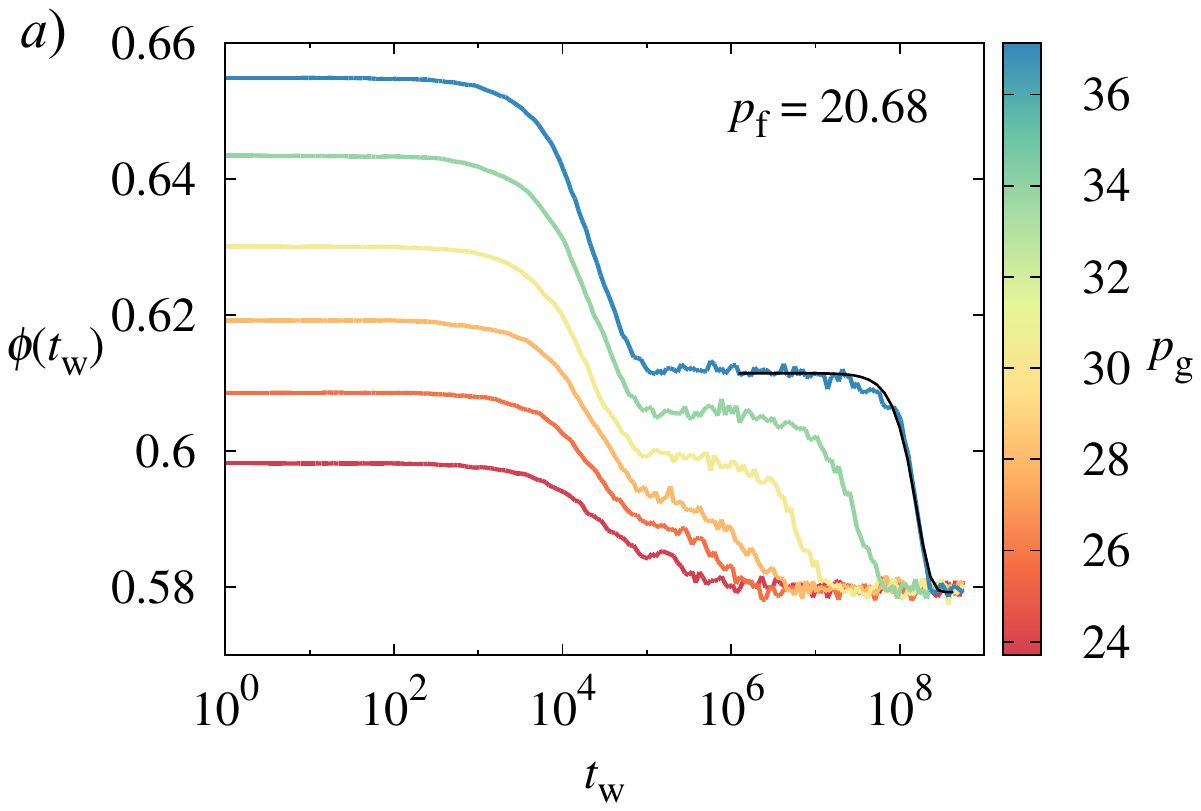}
\onefigure[width = 8.5cm]{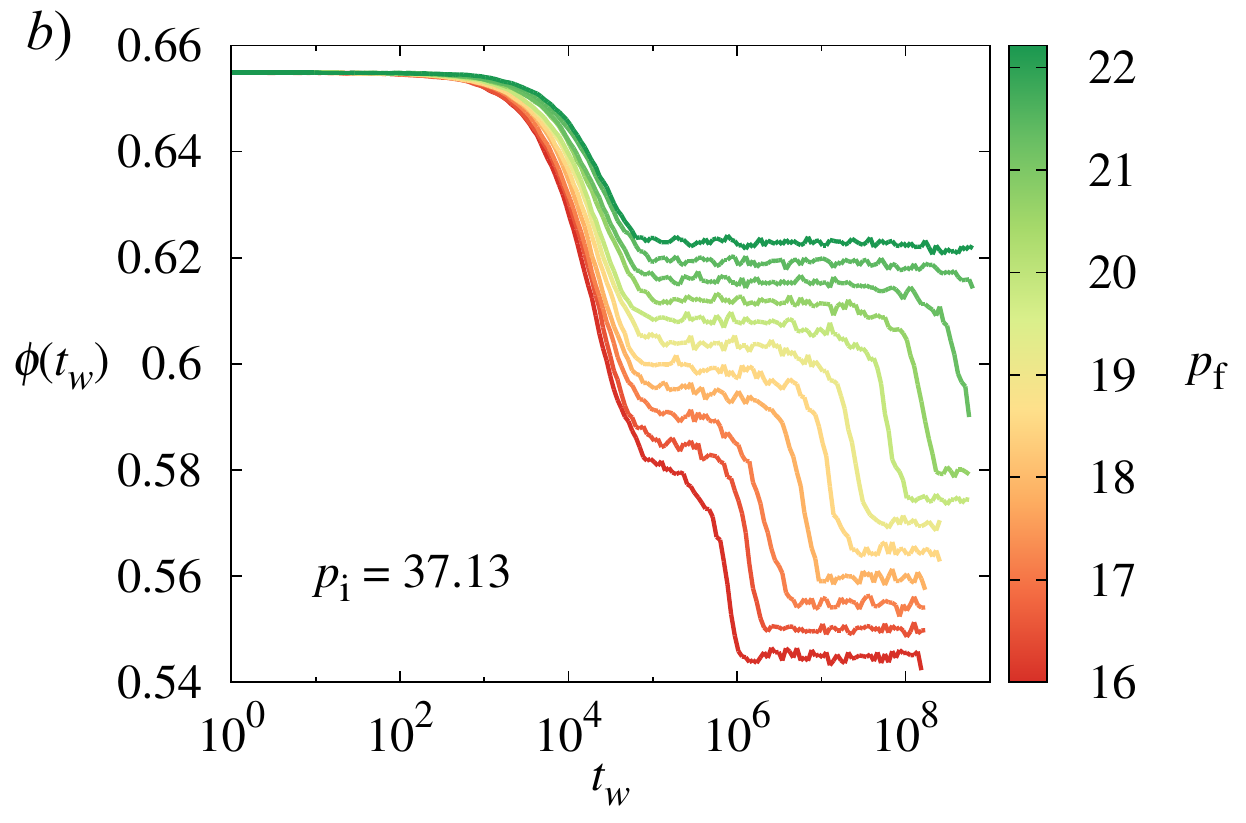}
\caption{a) Evolution of the packing fraction during 
melting from various glasses to the same fluid state at $p_f = 20.68$. 
The more stable a glass is, the larger the density difference 
between the glass and the fluid, and thus
the slower the melting. The black line is a 
fit to a compressed exponential decay with exponent 
$\beta = 2.66$. 
b) The same as a) but for a given initial glass prepared at $p_g = 37.13$ and melted to
different fluid states.} 
\label{fig:sudden_melt_II}
\end{figure}

We now focus on isobaric melting.
The melting time depends both on the initial glass and 
final fluid states~\cite{viejo2016}. In  
Fig.~\ref{fig:sudden_melt_II}a, we follow the melting of 
glasses prepared at various initial states to the same 
final fluid state, by measuring the dependence of the packing
fraction on the waiting time $\tw$ since the pressure was 
suddenly changed from $p_g$ to $p_f$. 
For each glass, we observe first a 
rapid expansion towards an intermediate density, during which the glass
structure is essentially unchanged. This corresponds
to following the nonequilibrium equation of state
in Fig.~\ref{fig:sudden_melt_I}a. This is followed 
by a second, much slower, expansion where the glass 
melts. As the density difference between the glass and the fluid increases,
the melting becomes much slower. Since these glasses melt to
the same fluid state, the 3 orders of magnitude increase 
in the melting times in Fig.~\ref{fig:sudden_melt_II}a directly 
translates into a similar growth of the stability ratio.  
In Fig.~$\ref{fig:sudden_melt_II}$b, we show how the same 
initial glass state melts into different fluid states.  
The nonequilibrium equation of state is the same in each case, but the 
intermediate density after the rapid expansion varies.
The higher the final pressure $p_f$ the longer the 
melting time, but since the fluid relaxation time changes as well, 
the stability ratio cannot directly be inferred from these plots. 

\begin{figure}
\onefigure[width = 8.5cm]{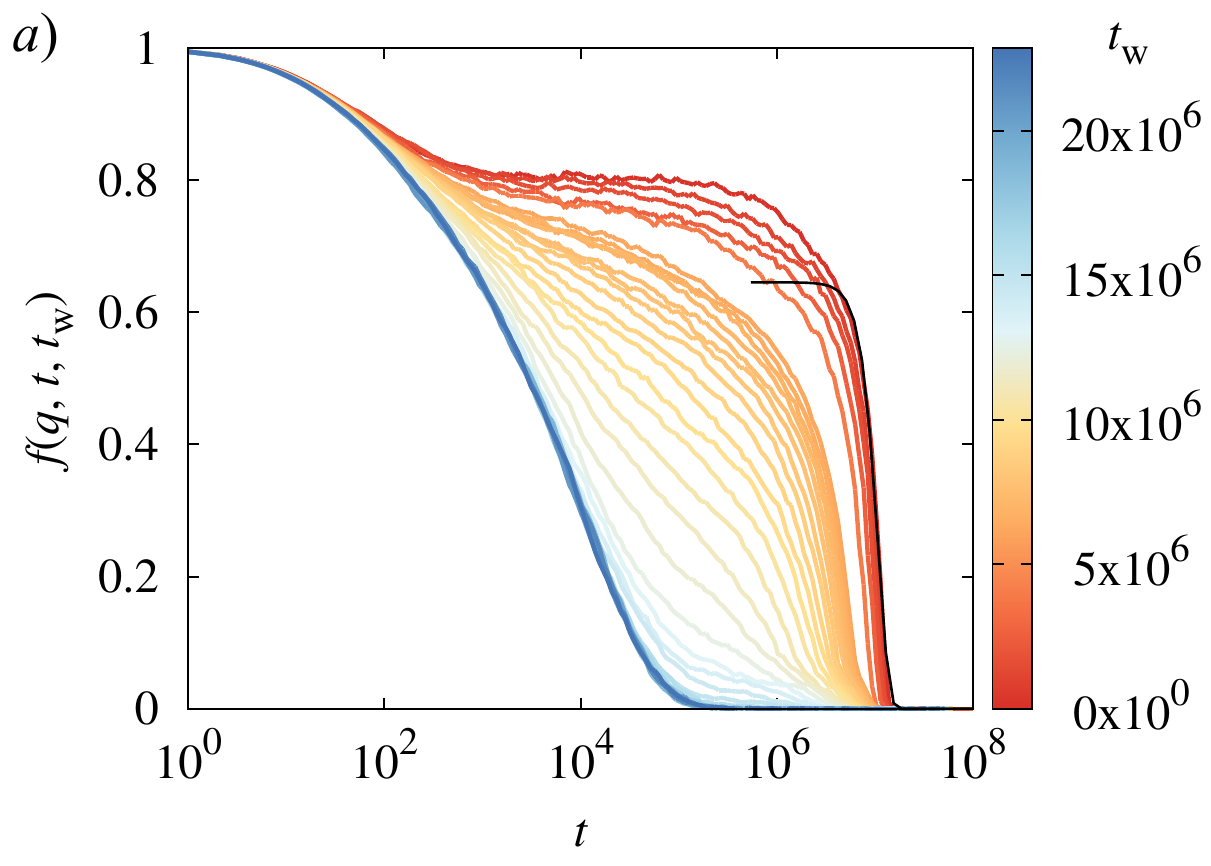}
\onefigure[width = 8.5cm]{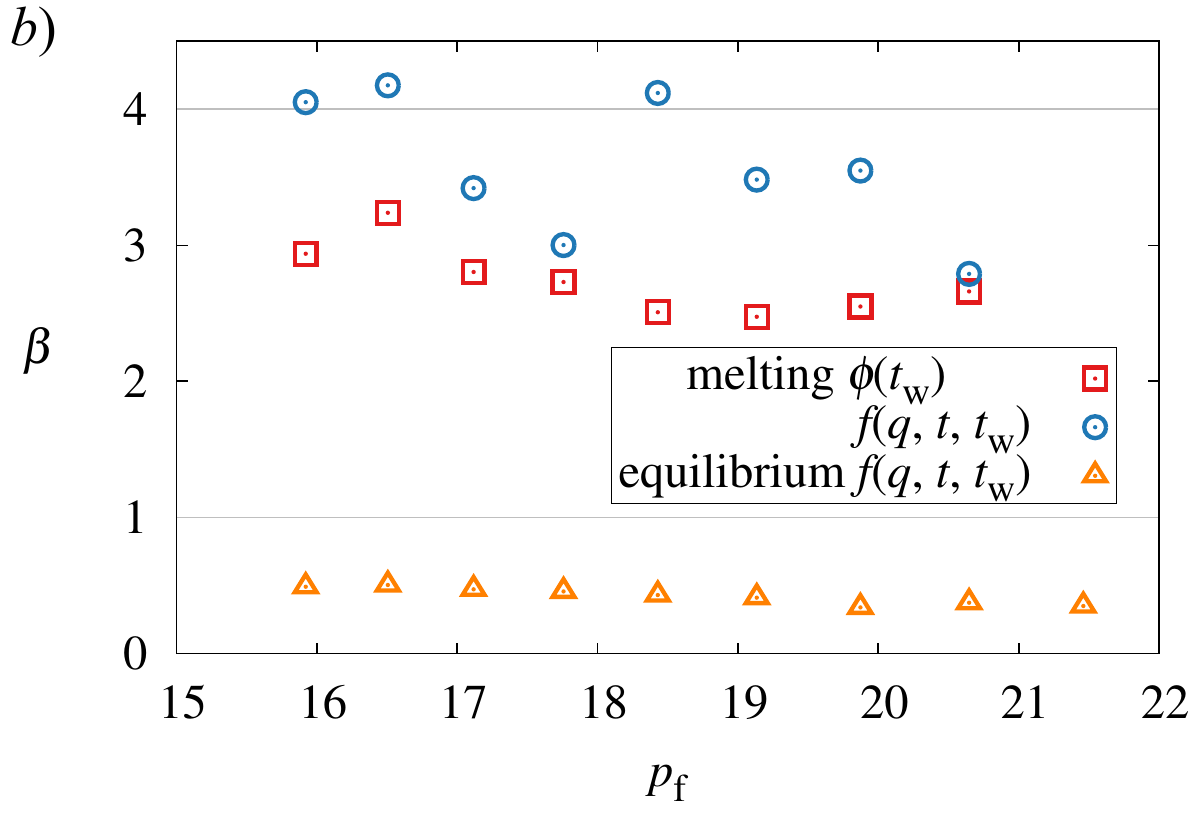}
\caption{a) Incoherent scattering function calculated as a stable glass prepared at $p_g = 37.13$ melts
at a constant pressure of $p_f = 18.50$ for a wide range of waiting times. 
The black line is a fit to a compressed exponential with 
exponent $\beta = 4.12$.
b) The exponent $\beta$ from the incoherent scattering functions
and from the packing fraction
during the melting of a stable glass prepared at $p_g = 37.13$ at various $p_f$.
The large values of $\beta$ during melting 
suggest that it proceeds by nucleation and growth of liquid 
bubbles in the glass. This contrasts strongly with equilibrium 
relaxation characterized by 
$\beta \approx 0.55$.
The horizontal lines mark $\beta = 4$ (homogeneous Avrami melting) 
and $\beta = 1$ (pure exponential decay).}
\label{fig:KWW}
\end{figure}

{\it Kinetics of melting ---}
Glasses produced by vapor deposition are films for which the fluid first 
appears at the free surface and invades the glass as 
a propagating front~\cite{swallen_2009}.
The melting time increases linearly with the film thickness 
until a crossover thickness
above which bulk-driven melting dominates~\cite{kearns_2010}. 
It has been suggested that 
bulk melting proceeds by the nucleation and growth of liquid 
bubbles~\cite{kearns_2010,Jack_2016}.
The crossover thickness then defines a characteristic length scale 
associated with bulk melting, which can become very large. 
Melting driven by the nucleation and growth of liquid bubbles is
described by Avrami kinetics~\cite{avrami_1939},
in which a time-dependent quantity $F(t)$
measured during melting will have a compressed exponential form, 
$F(t) = F_0 \exp[-(t/t_0)^\beta]$, with an exponent $\beta > 1$. For homogeneous 
nucleation in three dimensions, $\beta = 4$~\cite{avrami_1939}.
This analysis was applied to experimental vapor-deposited 
glasses~\cite{dawson_2012} and model spin 
systems~\cite{Jack_2016,gutierrez_2016}.

We consider the packing fraction of the system as a function of waiting time 
$\phi(\tw)$, 
and the incoherent scattering function $f(\vec{q}, t,\tw) = 1/N\sum_j 
\exp[i\vec{q}.(\vec{r}_j(t+\tw) - \vec{r}_j(\tw))]$ during melting.
Here $\vec{r}_j(t)$ is the position of particle $j$ at time $t$ and $\vec{q}$ is 
the wavevector of the first peak of the structure factor.  
The behaviour of $\phi(\tw)$ is shown in Fig.~\ref{fig:sudden_melt_II}, 
and that of $f(\vec{q}, t,\tw)$ in Fig.~\ref{fig:KWW}a.
The incoherent scattering function ages during melting. It displays a clear 
plateau and a slow, compressed decay for short $\tw$ which 
accelerates and becomes more stretched at long $\tw$, as expected 
for the transformation of a stable glass into an equilibrium fluid.

We extract the exponent $\beta$ for $\tw=0$ 
(for the melting process) and $\tw=\infty$
(for equilibrium). We also fit the long-time decay 
of $\phi(\tw)$ and get an independent estimate 
of $\beta$ for the melting. Example fits are shown as solid black lines in 
Figs.~\ref{fig:sudden_melt_II}a and \ref{fig:KWW}a.
The results for $\beta$ are compiled in Fig.~\ref{fig:KWW}b.
For $f(\vec{q}, t,\tw)$ we find $2.8 < \beta < 4.2$ and 
for $\phi(\tw)$, $2.5 < \beta < 3.2$. 
For comparison we get $\beta \sim 0.55$ at equilibrium. 
Clearly melting is well-described by compressed exponential functions, indicating that it starts slowly (nucleation) 
and then accelerates (growth), as in Avrami kinetics.
As $p_f$ increases towards $p_g$, $\beta$ should smoothly crossover to its equilibrium value, but this regime is outside the range shown in Fig.~\ref{fig:KWW}b.

However, we find $\beta < 4$, so it is likely our system 
deviates from the pure process with homogeneous nucleation.
In a spin model of melting by nucleation and growth, it was observed that in 
processes where nucleation was fast compared to growth, the 
exponent $\beta$ associated with melting was less than the 
Avrami prediction~\cite{gutierrez_2016}.
This may be the case in our system.  
If the nucleation process is inhomogeneous, we would also find $\beta < 4$.
If the local structure of the glass is correlated with its 
dynamics~\cite{wolynes_2009,Jack_2016}, melting would 
preferentially start from structurally disordered sites. 
To test this hypothesis, we melted 
the same initial glass configuration multiple times using independent dynamic 
trajectories to see if melting always begins in the same regions of the system. 
Although qualitative at this stage, our observations indicate that this
is the case, as we indeed find some regions where melting 
systematically begins very early on. However, we also found 
regions where melting begins only in some of the trajectories.
We plan to analyse these results more quantitatively,
in order to understand better the seeds of the melting process. 

\begin{figure}
\onefigure[width = 8.5cm]{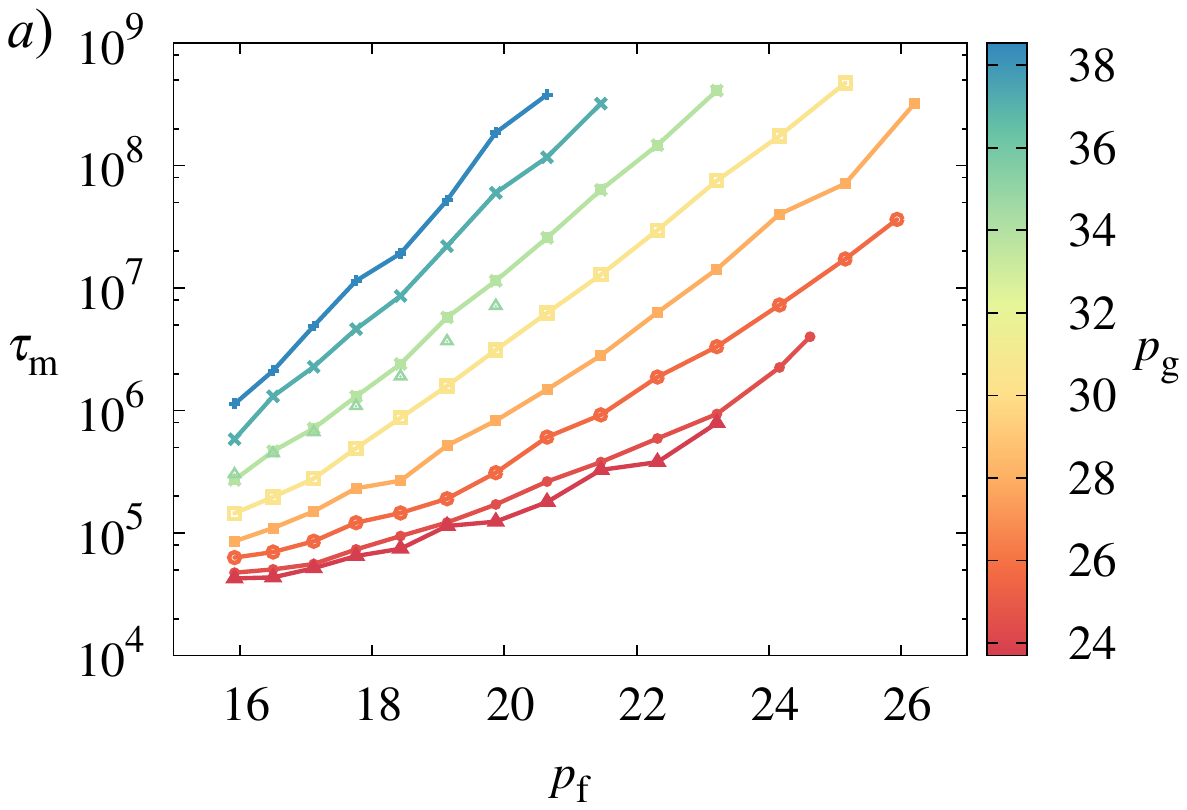}
\onefigure[width = 8.5cm]{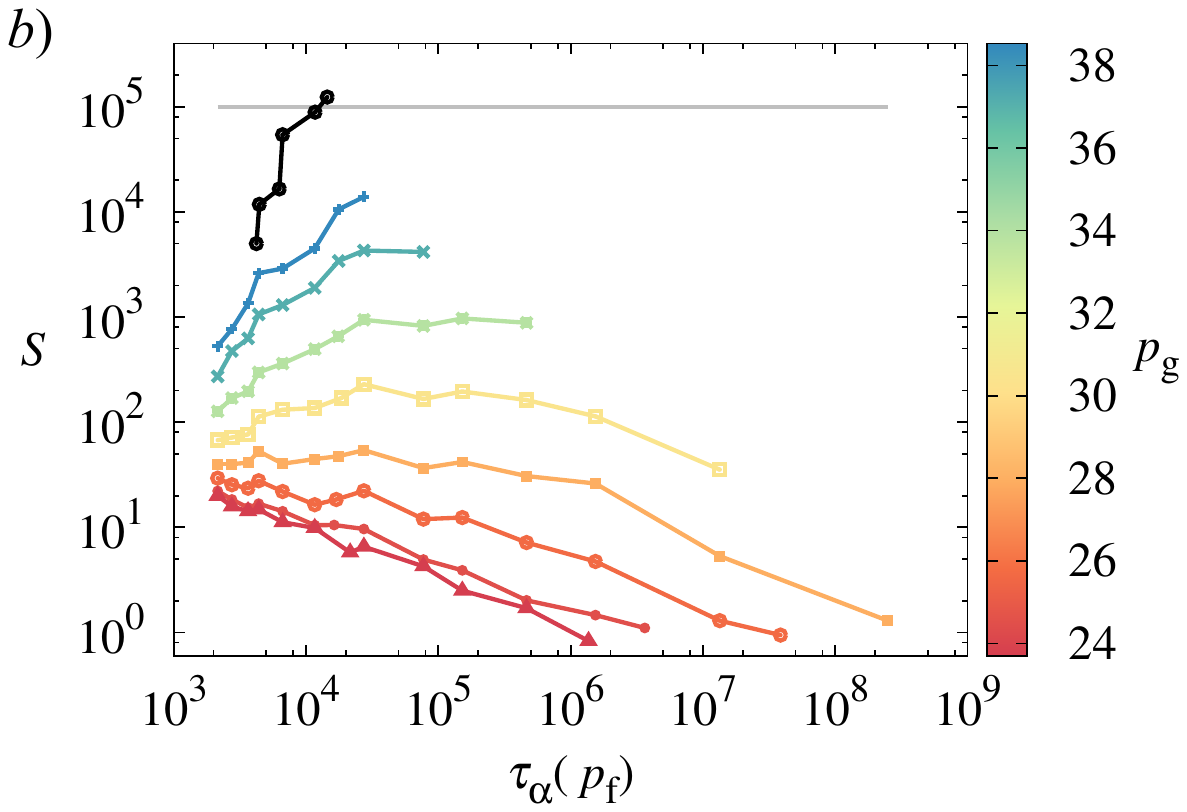}
\caption{a) Evolution of the melting time $\tm$ with $p_f$
for a range of stable glasses with $N=1000$ prepared at different 
initial pressures $p_i$. Additional symbols 
are for $N=8000$. 
b) Evolution of the stability ratio ${\cal S}$ with 
$\tau_\alpha(p_f)$ for the same glasses as in a). An additional black 
line shows the most stable glass we have produced so far, a system 
with $N=300$ and non-additive interactions. The most stable glass was prepared at $p_g = 42.90$ and ${\cal S}$ was highest when it was melted at $p_f = 18.46$.
The grey line marks ${\cal S}  = 10^5$, a typical stability for the most stable experimental vapor-deposited glasses.}
\label{fig:melting_time_const_p}
\end{figure}

We extract the melting time $\tm$ and equilibrium relaxation 
time $\tr$ from the decay of 
time correlation functions, namely $f(\vec{q}, t = \tm ,\tw=0) = 
f(\vec{q}, t = \tr ,\tw=\infty) = e^{-1}$. 
In Fig.~\ref{fig:melting_time_const_p}a we show the 
evolution of $\tm$ with $p_f$ for a range of 
stable glasses prepared at various $p_g$.  
Most systems have $N = 1000$ particles, but we also include 
measurements with $N=8000$ that show results consistent with 
the smaller systems. 
Increasing stability is observed by comparing the melting of 
different glasses at the same $p_f$. Glasses with 
higher $p_g$ have longer melting times and are thus more stable.

We finally consider the stability ratio, ${\cal S} = \tm / \tr$,
which has two trivial limits. It is close to unity 
both when the melting is performed at low pressures outside
the glassy regime, or at large pressures when $p_f \to p_g$. 
Therefore we expect ${\cal S}$ to display a maximum
at intermediate pressures, as confirmed 
in Fig.~\ref{fig:melting_time_const_p}b, which converts the data 
of Fig.~\ref{fig:melting_time_const_p}a into stability ratios. 
We use $\tr(p_f)$ (instead of $p_f$ itself)
for the horizontal axis, 
as this allows different systems to be compared on the same graph. 
Limitations on simulated timescales prevent us from being able to measure a
maximum for all glasses.
The largest ${\cal S}$ value we measure is 
${\cal S} = 10^{4.1}$, for a glass prepared 
at $p_g = 38.5$ and melted at $p_f = 20.6$.
On the same graph we show additional
results for the non-additive hard sphere system 
with $\epsilon = 0.2$ and $N = 300$ as a black line. 
For $p_g = 42.90$ and $p_f = 18.46$ 
we measure ${\cal S} = 10^{5}$, which is the largest 
stability ratio yet measured in a simulated bulk glass-former and 
is comparable to that of the most stable experimental vapor-deposited films.
By contrast, when we melt the same glasses at constant volume 
we again measure a maximum stability of ${\cal S} \approx 10^{2}$.
This confirms further that the high density of 
stable glasses is the key stabilising factor
against melting into a lower density fluid.

The fact that vapor deposited glasses are thin films and that the most stable of them are (presumably) out of equilibrium does not affect our conclusions.
The behaviour of our glasses is representative of films thick enough to melt by bulk processes. 
The degree of thermalisation at a given $p_g$ primarily controls the stability ratio. It does not matter that our states are thermalised rather than being slightly out of equilibrium: we will just measure a higher ${\cal S}$.

\begin{figure}
\onefigure[width = 8.5cm]{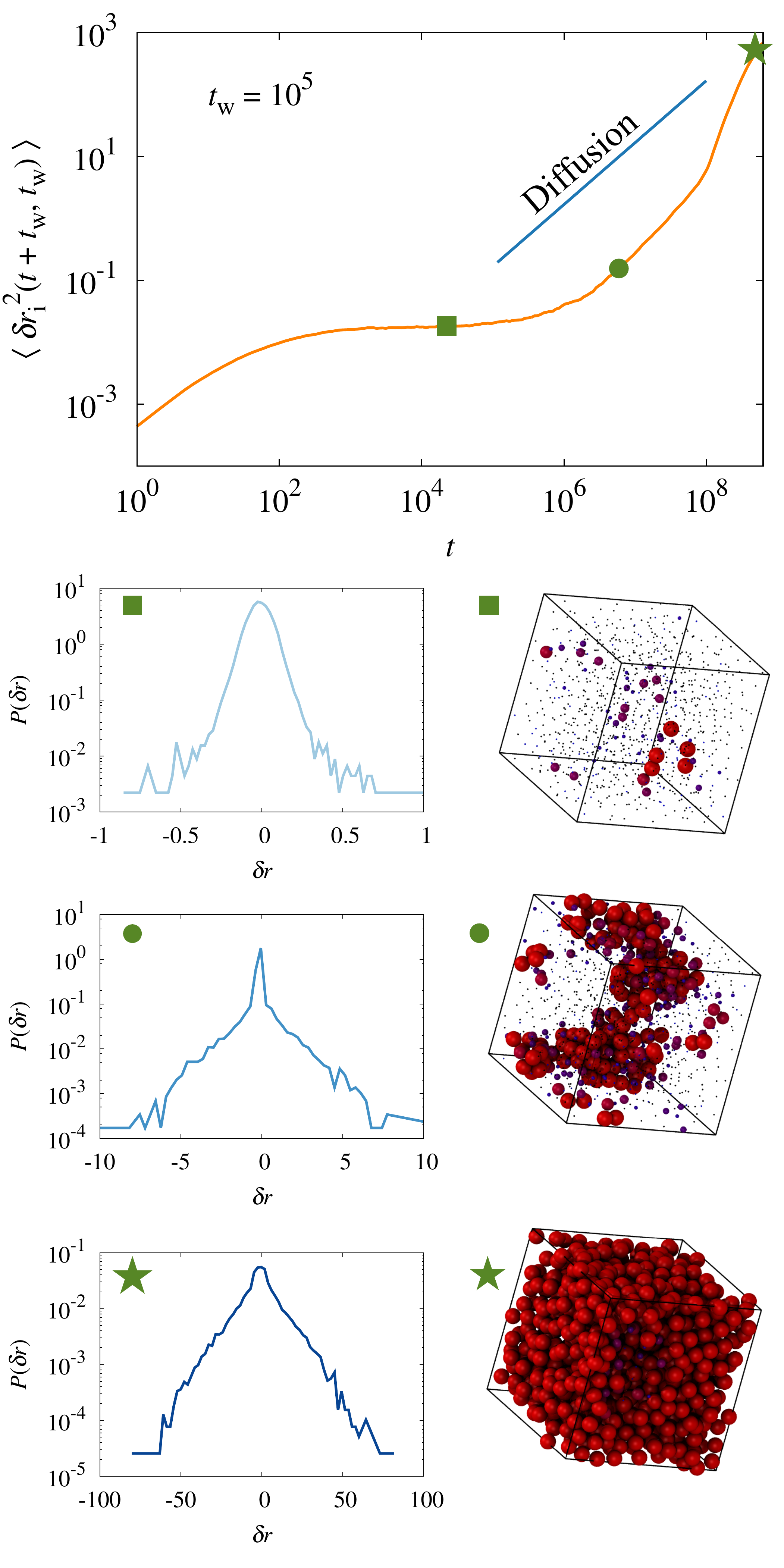}
\caption{The mean-squared displacement measured as a stable glass 
prepared at $p_g = 33.90$ melts at $p_f = 21.44$.
During melting, the mean-squared displacement appears faster
than diffusion (blue line).
Bottom panels show the van-Hove funciton for three 
times during the melting indicated by the symbols.
Squares: nucleation of liquid bubbles. 
Circles: growth of the liquid.
Stars: approach to equilibrium.
Snapshots show the most mobile particles 
defined by $\delta r > 0.22$.}
\label{fig:van_hove}
\end{figure}

{\it Microscopic view of melting ---}
To characterize melting in space and time we measure 
single particle mean-squared displacements~\cite{hocky_2014},
$\delta r_i^2(t+\tw,\tw)$. 
The corresponding probability distribution function is the van-Hove function,
$P(\delta r)$. 
If stable glasses melt through the nucleation and growth of liquid bubbles, we 
expect to see evidence in the van-Hove function that a sub-population of 
particles has melted while the other particles remain immobile.
We also expect to see evidence in measures of local mobility. 

In Fig.~\ref{fig:van_hove}, we show 
a typical mean-squared displacement measured
during melting. For three representative times 
we also show both the corresponding van-Hove distribution and a snapshot
of the system highlighting regions of large mobility.
The mean-squared displacement reaches a plateau corresponding
to localised particle motion in the glass before melting. 
When melting proceeds, there is a sudden upturn of the mean-squared 
displacement that appears faster than diffusion 
and corresponds to the melting process.
This fast increase is consistent with the growth phase of the Avrami picture. It can be very simply interpreted as a delayed onset of particle motion.
At very large times, diffusive behaviour will set in. 

Resolving this average behaviour in space and time, we observe 
that at early times before melting (squares), 
particles are trapped by their neighbours.
The van-Hove function takes the form of a time-independent gaussian 
distribution and particle mobility is low throughout the system.
As melting begins (circles), regions of high mobility 
appear, corresponding to the liquid bubbles.
The particles are divided into mobile and 
immobile populations, so the van-Hove function appears as a superposition of two 
distributions. When melting has finished (stars), most particles have 
moved far from their initial positions and the van-Hove function takes the 
form of a gaussian distribution whose width grows linearly with time.

In snapshots of the system during melting, 
the length scale associated with the size and separation 
of mobile regions does not appear to be as large as 
the one inferred from the crossover to bulk melting in experiments
on ultrastable glass films~\cite{Jack_2016, kearns_2010}.
Our simulations of a larger system with $N=8000$
confirm that finite size effects are small, 
and do not indicate that the dynamic melting 
length scale becomes larger in larger systems.
A possible explanation is that the time for liquid regions to 
nucleate and the time for them to grow respond in different ways to 
the density difference. For more stable glasses, 
the density difference is larger and the time 
for liquid regions to nucleate and grow should both increase.
If the time for growth increases faster than the time for nucleation then the 
size and separation of the regions (and the associated length scale) should 
be small~\cite{Jack_2016}. 

{\it Conclusion ---}
It was recently claimed that glass configurations
prepared using the swap Monte Carlo method closed 
the large timescale gap between ordinary simulations
and experiments~\cite{NBC17}.
In this work, we have demonstrated that these configurations
correspond to bulk glasses that are indeed `ultrastable'~\cite{swallen_2007}.
For the best configurations produced with this 
technique, we have measured 
values of the stability ratio comparable to those obtained
for the most stable glass films produced experimentally using vapor deposition.
The reasons for this large increase in kinetic stability are twofold.  
Firstly, the swap Monte Carlo algorithm allows systems to be 
equilibrated at previously unreachable low temperatures and high packing 
fractions. Secondly, these glasses are stabilised by melting them at 
constant pressure, as in experiments. We found 
that a large density difference between the glass and the liquid
considerably slows down the melting process and
presented microscopic evidence that bulk melting
proceeds through the nucleation and growth 
of liquid bubbles inside the bulk glass.

The melting process deserves further exploration, especially as it is 
tied to deeper issues about the nature of the glass 
transition~\cite{Jack_2016,berthier_2011_rev}.
There are several open questions regarding 
spatio-temporal aspects 
of the melting process which can be answered following the 
approach proposed in the present work. We plan to investigate 
different types of glass-formers, using for instance 
Lennard-Jones interactions, to reproduce 
more quantitatively the thermodynamics of real ultrastable materials. 
We need to improve our simulation tools to investigate 
larger systems so we can understand the length scales
associated wth melting as well the possible existence and nature of
sites where melting is initiated preferentially.

\acknowledgements
We thank R. Jack for useful discussions. The research leading to
these results has received funding from the European Research
Council under the European Union's Seventh Framework 
Programme (No. FP7/2007-2013)/ERC Grant Agreement No.306845, and was supported by a grant from the Simons Foundation (No. 454933, L. Berthier).

\bibliographystyle{eplbib}
\bibliography{stable_glass_EPL}

\end{document}